\documentstyle[12pt,epsfig] {article}
\begin {document}
\begin{center}
\vskip 1.5 truecm
\begin{Large}
{\bf Role of the non-resonant} {\bf background in the}
$\rho^0${\bf -meson diffractive electro- and photoproproduction.}\\
\end{Large}
\vspace{.5cm}
M.G.Ryskin and Yu.M.Shabelski \\
\vspace{.5cm}
Petersburg Nuclear Physics
Institute, \\ Gatchina, St.Petersburg 188350 Russia \\
\end{center}
\vspace{1cm}
\begin{abstract}
The background due to the direct diffractive dissociation of the photon
into the $\pi^+\pi^-$-pair to the "elastic" diffractive $\rho^0$-meson
production in electron-proton collisions is calculated.
The amplitude for the background process $\gamma p \rightarrow \pi^+ \pi^-
p$ is proportional to the $\pi$-meson - proton cross section. Therefore,
describing the HERA data, we can estimate $\sigma (\pi p)$ at energy
$s_{\pi p}\sim (2\,-\,3)\cdot 10^{3}$ GeV$^2$ that is considerably higher
the existing data. At large $Q^2$ the interference between resonant and
non-resonant $\pi^+ \pi^-$ production leads also to the more slow increase
of the $\sigma^L/\sigma^T$ ratio with the mass of the $2\pi$ (i.e.
$\rho^0$-meson) state.
\end{abstract}
\vspace{3cm}

E-mail: $\;$ RYSKIN@thd.PNPI.SPB.RU  \\

E-mail: $\;$ SHABELSK@vxdesy.desy.de \\

\newpage

\section{Introduction}
It was noted many years ago that the form of the $\rho$-meson peak is
distorted by the interference between resonant and non-resonant
$\pi^+\pi^-$ production. For the case of "elastic" $\rho^0$
photoproduction the effect was studied by P.S\"oding in \cite{so} and
S.Drell \cite{dr} (who considered the possibility to produce the pion
beam via the $\gamma\to \pi^+\pi^-$ process).  At high energies the main
(and the only) source of background is the Drell-Hiida-Deck process \cite{d}
(see fig. 1). The incoming photon fluctuates into the pion pair and then
$\pi p$-elastic scattering takes place. Thus the amplitude for the
background may be written in terms of the pion-proton cross section.
Recently the difractive elastic production of $\rho^0$-mesons was
measured at HERA \cite{zeus,zeus1,H1,H11} both for the cases of
photoproduction i.e. $Q^2 = 0$ and of $Q^2 \geq 4$ GeV$^2$ (the so
called deep inelastic scattering, DIS, regime). It was demonstrated
\cite{zeus,H1} that the interference with some non-resonant background
is indeed needed to describe the distribution over the mass - $M$ of
$\pi^+\pi^-$ pair.

It was proposed by M.Arneodo that this effect can be used to estimate
the value of $\sigma_{\pi p}$ from HERA data at high energies \cite{ar}
($\sqrt{s}=W\sim 40-55$ GeV), in the range which is not otherwise
acceptable.

In Sect. 2 the formulae for the $2\pi$ background which are valid for
the DIS as well as for the photoproduction region are presented. The
expression differs slightly from the S\"oding's one as we take into
account the pion form factor and the fact that one pion propagator is
off-mass shell. We consider also the absorbtion correction comming from
the diagram where both pions ( $\pi^+$ and $\pi^-$) directly interact
with the target proton. The role of the interference in photoproduction
is discussed in Sect. 3, where we estimate the $\pi$-proton cross section
$\sigma_{\pi p}\simeq 30$ mb at $\sqrt{s}\sim 50$ GeV. Finally in Sect. 4
we compute the amplitude for a pion pair production in DIS. At large
$Q^2\sim 10\,-\,30$ GeV$^2$ the background amplitude becomes relatively
small, but still not negligible. It changes the ratio of the longitudinal
to transverse $\rho$-meson production cross section and leads to the more
slow increase of $R=\sigma^L/\sigma^T$ with $M^2$.

\section{Production amplitudes}

The cross section of $\rho^0$ photo- and electroproduction may be written
as:
\begin{equation}
\frac{d\sigma^D}{dM^2dt}\; =\; \int d\Omega
|A_{\rho}+A_{n.r.}|^2 \;,
\end{equation}
where $A_{\rho}$ and $A_{n.r.}$ are the resonant and non-resonant parts of
the production amplitude, $D=L\, ,T$ for longitudinal and transverse photons,
$t = $ -{\bf q}$_t^2$ is the momentum transfered to the proton and
$d\Omega =d\phi dcos(\theta)$, where $\phi$ and $\theta$ are the azimutal
and polar angles between the $\pi^+$ and the proton direction in the
$2\pi$ rest frame.

\subsection{Amplitude for resonant production}

The dynamics of vector meson photo- and electroproduction was discussed
in the framework of QCD in many papers (see, e.g. [10-13]). However here
we will use the simple phenomenological parametrization of the production
amplitude because our main aim is the discussion of the interference
between resonant and non-resonant contributions. So the amplitude for
resonant process $\gamma p \to \rho^0 p$; $\rho^0 \to \pi^+ pi^-$ reads:
\begin{equation}
A_{\rho}\; =\; \sqrt{\sigma_{\rho}}e^{-b_\rho q^2_t/2}\frac{\sqrt{M_0\Gamma}}
{M^2-M^2_0+iM_0\Gamma}\frac{H^D(\theta ,\phi)}{\sqrt{\pi}} \;.
\end{equation}
To take into account the phase space available for the $\rho\to\pi^+\pi^-$
decay we use the width $\Gamma=\Gamma_0\left(\frac{M^2-4m^2_\pi}
{M^2_0-4m^2_\pi}\right)^{3/2}$ (with $\Gamma_0=151$ MeV and $M_0=768$ MeV
-- its mass); $b_\rho$ is the $t$-slope of the "elastic" $\rho$ production
cross section $\sigma_\rho\equiv d\sigma(\gamma p \to \rho^0 p)/dt$
(at $t=0$) and the functions $H^D(\theta ,\phi),\; D = T, L$ describe the
angular distribution of the pions produced through the $\rho$-meson decay:
\begin{equation}
H^L\; =\; \sqrt{\frac 3{4\pi}}cos\theta \;,
\end{equation}
\begin{equation}
H^T\; =\; \sqrt{\frac 3{8\pi}}sin\theta\cdot e^{\pm i\phi} \;.
\end{equation}
Note that for transverse photons with polarization vector $\vec e$
one has to replace the last factor $e^{\pm i\phi}$ in eq. (4) by the
scalar product $(\vec e\cdot\vec n)$, where $\vec n$ is the unit vector
in the pion transverse momentum direction.

\subsection{Amplitude for non-resonant production}

The amplitude for the non-resonant process $\gamma p \to \pi^+\pi^-p$ is:
\begin{equation}
A_{n.r.}\; =\;\sigma_{\pi p}F_\pi (Q^2)e^{bt/2}\frac{\sqrt{\alpha}}
{\sqrt{16\pi^3}} B^D\sqrt{z(1-z)\left|\frac{dz}{dM^2}\right|
\left(\frac{M^2}4-m^2_\pi\right)|cos\theta|} \;,
\end{equation}
where $b$ is the $t$-slope of the elastic $\pi p$ cross section,
$F_{\pi} (Q^2)$ is the pion electromagnetic form factor
($Q^2=|Q^2_\gamma| > 0$ is the virtuality of the incoming photon),
$\alpha= 1/137$ is the electromagnetic coupling constant and $z$ -- the
photon momentum fraction carried by the $\pi^-$ -meson; $\sigma_{\pi p}$
is the total pion-proton cross section.

The factor $B^D$ is equal to
\begin{equation}
B^D\; =\; \frac{(e^D_\mu\cdot k_{\mu -})f(k'^2_-)}{z(1-z)Q^2+m^2_\pi+k^2_{t-}}
-\frac{(e^D_\mu\cdot k_{\mu +})f(k'^2_+)}{z(1-z)Q^2+m^2_\pi+k^2_{t+}}
\end{equation}
 For longitudinal photons the products $(e^L_\mu\cdot k_{\mu \pm})$ are:
 $(e^L_\mu\cdot k_{\mu -})=z\sqrt{Q^2}$ and
$(e^L_\mu\cdot k_{\mu +})=(1-z)\sqrt{Q^2}$, while for the transverse
photons we may put (after averaging) $e^T_\mu\cdot e^T_\nu =\frac
 12\delta^T_{\mu\nu}$.

Expressions (5) and (6) are the result of straitforward calculation of
the Feynman diagram fig. 1. The first term in (6) comes from the graph
fig. 1 (in which the Pomeron couples to the $\pi^+$) and the second one
reflects the contribution originated by the $\pi^- p$ interaction.  The
negative sign of $\pi^-$ electric charge leads to the minus sign of the
second term. We omit here the phases of the amplitudes. In fact, the
common phase is inessential for the cross section, and we assume that the
relative phase between $A_{\rho}$ and $A_{n.r.}$ is small (equal to zero)
as in both cases the phase is generated by the same 'Pomeron'
\footnote{Better to say -- 'vacuum singularity'.} exchange.

The form factor $f(k'^2)$ is written to account for the virtuality
($k'^2\neq m^2_\pi$) of the t-channel (vertical in fig. 1) pion. As in
fig. 1 we do not deal with pure elastic pion-proton scattering, the
amplitude may be slightly suppressed by the fact that the incoming pion
is off-mass shell. To estimate this suppression we include the form
factor (chosen in the pole form)
\begin{equation}
f(k'^2)=1/(1 + k'^2/m'^2)
\end{equation}
The same pole form was used for $F_\pi (Q^2)=1/(1 + Q^2/m^2_\rho)$.
In the last case the parameter $m_\rho = M_0$ is the mass of the
$\rho$-meson -- the first resonance on the $\rho$-meson (i.e. photon)
Regge trajectory, but the value of $m'$ (in $f(k'^2)$) is expected to be
larger. It should be of the order of mass of the next resonance from
the Regge $\pi$-meson trajectory; i.e. it should be the mass of $\pi
(1300)$ or $b_1 (1235)$. Thus we put $m'^2=1.5$ GeV$^2$.

Finaly we have to define $k'^2_\pm$ and $k_{t\pm}$.
\begin{equation}
\vec k_{t-}=-\vec K_t+z\vec q_t\;\;\;\;\;\;\;
\vec k_{t+}=\vec K_t+(1-z)\vec q_t
\end{equation}
and
\begin{equation}
k'^2_-=\frac{z(1-z)Q^2+m^2_\pi+k^2_{t-}}{z},\;\;\;\;\;\;
k'^2_+=\frac{z(1-z)Q^2+m^2_\pi+k^2_{t+}}{1-z} \;.
\end{equation}
In these notations
$$M^2=\frac{K^2_t+m^2_\pi}{z(1-z)},\;\;\;\;
\;\;\;\; dM^2/dz=(2z-1)\frac{K^2_t+m^2_\pi}{z^2(1-z)^2}$$
and $ z=\frac 12\pm\sqrt{1/4-(K^2_t+m^2_\pi)/M^2}$ with the pion
transverse (with respect to the proton direction) momentum $\vec K_t$
(in the $2\pi$ rest frame) given by expression
$K^2_t=(M^2/4 - m^2_\pi)sin^2\theta$. Note that the positive values of
$cos\theta$ correspond to $z \geq 1/2$ while the negative ones
$cos\theta < 0$ correspond to $z \leq 1/2$.

\subsection{Absorptive correction}

To account for the screening correction we have to consider the diagram
fig. 2, where both pions interact directly with the target. Note that all
the rescatterings of one pion (say $\pi^+$ in fig. 1) are already included
into the $\pi p$ elastic amplitude. The result may be written in form of
eq. (5) with the new factor $\tilde{B}^D$ instead of the old one
$B^D=B^D(\vec K_t,\vec q)$:
\begin{equation}
\tilde{B}^D\; = \; B^D(\vec K_t,\vec q)-\int C\frac{\sigma_{\pi p}
e^{-bl^2_t}} {16\pi^2}B^D(\vec K_t-z\vec l_t,\vec q)d^2l_t
\end{equation}
where the second term is the absorptive correction (fig. 2) and $l_\mu$
is the momentum transfered along the 'Pomeron' loop. The factor $C>1$
reflects the contribution of the enhacement graphs with the diffractive
exitation of the target proton in intermediate state. In accordance with
the HERA data \cite{H2}, where the cross section of "inelastic" (i.e.
with the proton diffracted) $\rho$ photoproduction was estimated as
$\sigma^{inel}\simeq 0.5 \sigma^{el}$ we choose $C=1.5 \pm 0.2$.

\section{Photoproduction cross section}
The results at $Q^2=0$ (photoproduction) are shown in fig.3. Both slope
parameters, $b_{\rho}$ in eq.(2) and $b$ in eq.(5) were assumed to be
equal to 10 GeV$^{-2}$. The solid curve corresponds to the resonant
production contribution. It is normalized to the experimental ZEUS data
\cite{zeus} near the $\rho$ peak\footnote{The H1 Coll. cross section
\cite {H1} near the $\rho$ peak is about 1.5 times smaller than ZEUS Coll.
cross section \cite{zeus}, so we can not describe both experiments
simultaneously.}. The contributions of non-resonant production and its
interference with resonant one depend on the form factor $f(k'^2)$, the
screening corrections and the value of $\sigma_{\pi p}$.  To demonstrate
their role we present in fig.3 six variants of non-resonant (dashed
curves) and interference (dotted curves) contributions. The upper dashed
curve at the right-hand-side of fig.3 corresponds to the simplest
calculation for $f(k'^2)\equiv 1$ and without screening correction.
Second (to the down) dashed curve shows the same calculation but with
screening correction.  Here the value of $\sigma_{\pi p} = 28$ mb was
used. Third and fourth curves show the results of calculations with form
factor $f(k'^2)$, eq.(7), and $\sigma_{\pi p} = 30$ mb, without and with
screening corrections, respectively. Fifth (which is practically coinside
with the fourth one) and sixth curves correspond to the calculations with
form factor $f(k'^2)$, eq.(7), $\sigma_{\pi p} = 28$ mb,  without and
with screening corrections.

For interference contribution the upper curve at left-hand-side of fig.3
presents the case $f(k'^2)\equiv 1$, $\sigma_{\pi p} = 28$ mb and without
screening correction.  Next two curves correspond to the calculations
with form factor,eq.(7), without screening correction and for
$\sigma_{\pi p} = 30$ mb and $\sigma_{\pi p} = 28$ mb, respectively. The
fourth curve shows the variant with $f(k'^2)\equiv 1$, $\sigma_{\pi p} =
28$ mb and with screening correction.  Fifth and sixth curves correspond
to the calculations with form factor $f(k'^2)$, eq.(7), with screening
corrrection and for $\sigma_{\pi p} = 30$ mb and $\sigma_{\pi p} = 28$
mb, respectively.

The sum of all three contributions for the same six variants for
background and interference terms are presented in fig.4. Solid and
dashed curves here show the calculations without and with screening
correction, respectively. The difference between the curves is as a rule
smaller than the experimental errors. One can see quite reasonable agreement
with ZEUS experimental data \cite{zeus} and some difference between the
considered variants that can allow one to choice the best variant if the
accuracy of the data will increase. It is necessary to note that the
background and interference contribution have the same sign at
$M_{\pi^+\pi^-} < M_{\rho}$ and different signs at
$M_{\pi^+\pi^-} > M_{\rho}$ where these contributions cancel each other
in part. So the last region is lesser sensitive to the value of
$\sigma_{\pi p}$. The to-day HERA data \cite{zeus} are collected from the
range of $\gamma p$ energy $W=60-80$ GeV. Mean energy of a pion, produced
directly by the photon, is $E_\pi\sim E_{\gamma}/2$. Thus the cross section
$\sigma_{\pi p}$ corresponds to $s_{\pi p}\sim W^2/2 \sim (1.8-3.2)\cdot
10^3$ GeV$^2$ and one can see that cross section $\sigma_{\pi p} = 28
\div 30$ mb (which is extrapolation by hand from lower energy region) is
in more or less reasonably agreement with the data. Of course, more
accurate analysis is needed for the extraction of $\sigma_{\pi p}$ with
error bars.

\section{Electroproduction cross section}

At very large $Q^2$ the background amplitude (5) becomes negligible as, even
without the additional form factor (i.e. at $f(k'^2)\equiv 1$), the
non-resonance cross section falls down as $1/Q^8$ \footnote{In the
amplitude $A_{n.r.}$ one factor $1/Q^2$ comes from the electromagnetic
form factor $F_\pi(Q^2)$ and another one -- from the pion propagator
(term - $z(1-z)Q^2$ in the denominator of $B^D$ (see eq.(6)).}, while
experimentally \cite{zeus1,H11} the $Q^2$ behaviour of the elastic $\rho$
cross sections has been found to be $1/Q^n$, with $n\sim 5 (<8!)$.

Nevertheless, numerically at $Q^2\sim 10\; GeV^2$ the background as well
as interference contributions are still important. The results of our
calculations of $(Q^2 + M_{\rho}^2)^2 d\sigma /dM$ with $\sigma_{\pi p}$ =
28 mb, with form factor, eq.(7), and with screening correction are
presented in fig.5. Dashed curve show the case of $\pi^+ \pi^-$ pair
photoproduction, four solid curves correspond to the electro production
at $Q^2$ = 1, 4, 10 and 30 GeV$^2$, the dotted curve to the value
$Q^2$ = 30 GeV$^2$ and with $f(k'^2)\equiv 1$. The dash-dotted curve
corresponds to only resonant contribution at $Q^2$ = 30 GeV$^2$ and with
form factor. This value of $(Q^2 + M_{\rho}^2)^2 d\sigma/dM$ should be
independent on $Q^2$ in the case of pure $\rho$-dominant model. Really
it has some slight $Q^2$ dependence.

The background as well as interference contributions lead to the
nontrivial behaviour of the ratio $R=\sigma^L/\sigma^T$ with the two pion
mass $M_{2\pi}=M$. In the theoretical formulae which we used the index
$D=L,T$ denotes the polarization of the incoming photon. On the other
hand experimentally one measures the $\rho$-meson polarization, fitting
the angular distribution of decay pions. To reproduce the procedure we
take the flows of initial longitudinal and transverse photons to be equal
to each other ($\epsilon =N^L/N^T=1$, which is close to HERA case) and
reanalyse the sum of cross sections ($\sigma =\sigma^L+\sigma^T$) in a
usual way, selecting the constant and the $cos^2\theta$ parts.
\begin{equation}
I_0\; =\; \int\limits^1_{-1}\sigma(\theta)dcos\theta\; ;\;\;\;\;\;
I_2\; =\; \int\limits^1_{-1}\sigma(\theta)\frac{15}4(3cos^2\theta -1)dcos\theta
\end{equation}
In these terms the density matrix element $r_{00}=(2I_2+3I_0)/9I_0$ and
\begin{equation}
R\; =\; \sigma^L/\sigma^T\; =\; \frac{r_{00}}{1-r_{00}}
\end{equation}
Namely these last ratios (12) are presented in fig.6 for different $Q^2$
values. One can see that they depend both on $Q^2$ and $M$. These ratios
increase with $M$ for $M > M_{\rho}$ and this increase becomes more weak
for large $Q^2$.

\section{Conclusion}
We presented simple formulae for the background to 'elastic' $\rho$-meson
photo- and electroproduction which account for the absorptive correction
and virtuality of the  pion. The role of $\rho$-meson -- background
interference is not negligible even at $Q^2\sim 10$ GeV$^2$,  especially
for the $\sigma^L/\sigma^T$ ratio. A reasonable description of the ZEUS
photoproduction data is obtained with a value of the $\pi p$ total cross
section $\sigma_{\pi p} = 28 \div 30$ mb at energy
$s_{\pi p}\sim (2-3)\cdot 10^3\; GeV^2$. We consider this as an indication
to the growth of $\pi p$ cross section with energy in the energy region
higher than the region of existing direct measurements.

We are grateful to M.Arneodo for stimulating discussions. The paper is
supported by INTAS grant 93-0079.

\newpage

\begin{center}
{\bf Figure captions}\\
\end{center}

Fig. 1. Feynman diagram for the two pion photo-(electro)production.

Fig. 2. Diagram for the absorbtive correction due to both pions
rescattering.

Fig. 3. Resonant (Breit-Wigner, solid curve), non-resonant background
(dashed curves) and their interference (dotted curves) contributions to
$\gamma p \to \pi^+ \pi^- p$ reaction. The two last contributions are
calculated with six different assumptions, see text.

Fig. 4. Distribution over the mass of two pions in photoproduction. Solid
and dashed curves correspond to the calculations without and with
screening correction, respectively. ZEUS Coll. data points \cite{zeus}
are presented also.

Fig. 5. $\pi^+ \pi-$ mass distribution in photo- (dashed curve 1) and
electroproduction at $Q^2 =$ 1 GeV$^2$ (solid curve 2), 4 GeV$^2$ (curve 3),
10 GeV$^2$ (curve 4) and 30 GeV$^2$ (curves 5) with (solid curve) and
without (dotted curve) form factor. The only resonant contribution at
$Q^2$ = 30 GeV$^2$ is shown by dash-dotted curve.

Fig. 6. The ratio $R=\sigma^L/\sigma^T$ in electroproduction process as a
function of pion pair mass at $Q^2 =$ 1 GeV$^2$ (curve 1), 10 GeV$^2$
(curve 2) and 30 GeV$^2$ (curves 3) with (solid
curve) and without (dashed curve) form factor.

\begin{figure}
\centerline{\epsfig{file=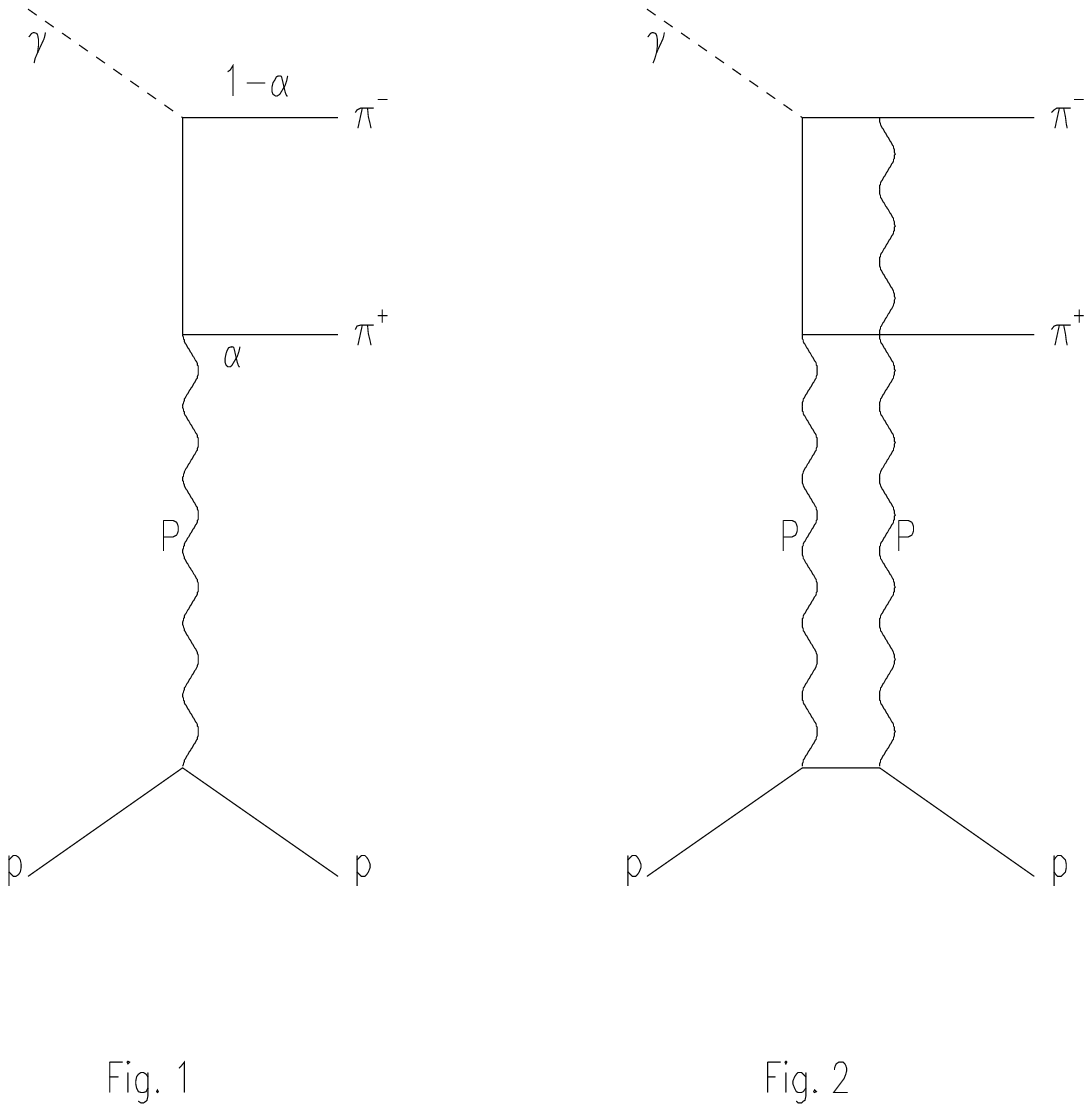,width=16cm}}
\end{figure}

\begin{figure}
\centerline{\epsfig{file=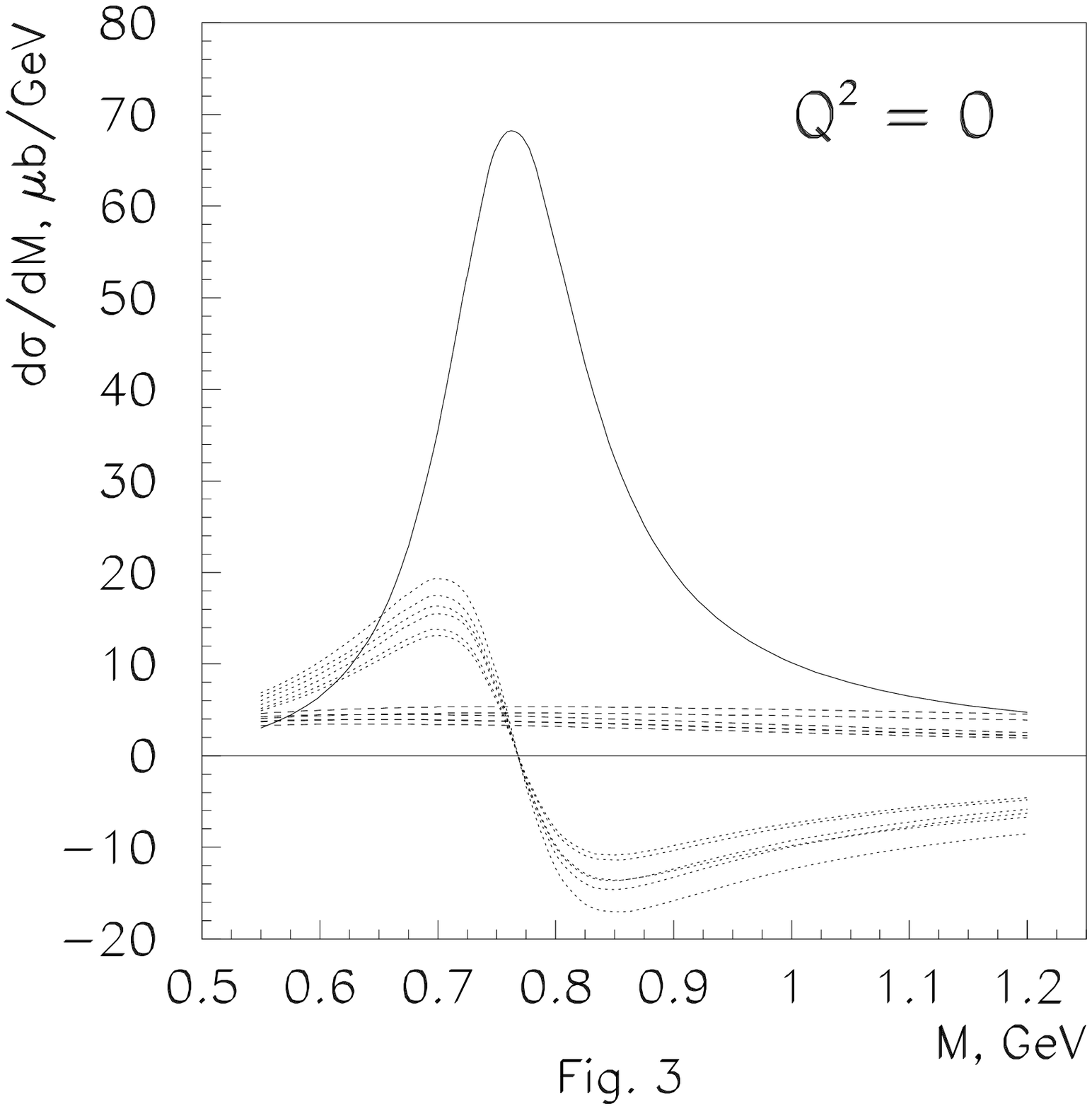,width=16cm}}
\end{figure}

\begin{figure}
\centerline{\epsfig{file=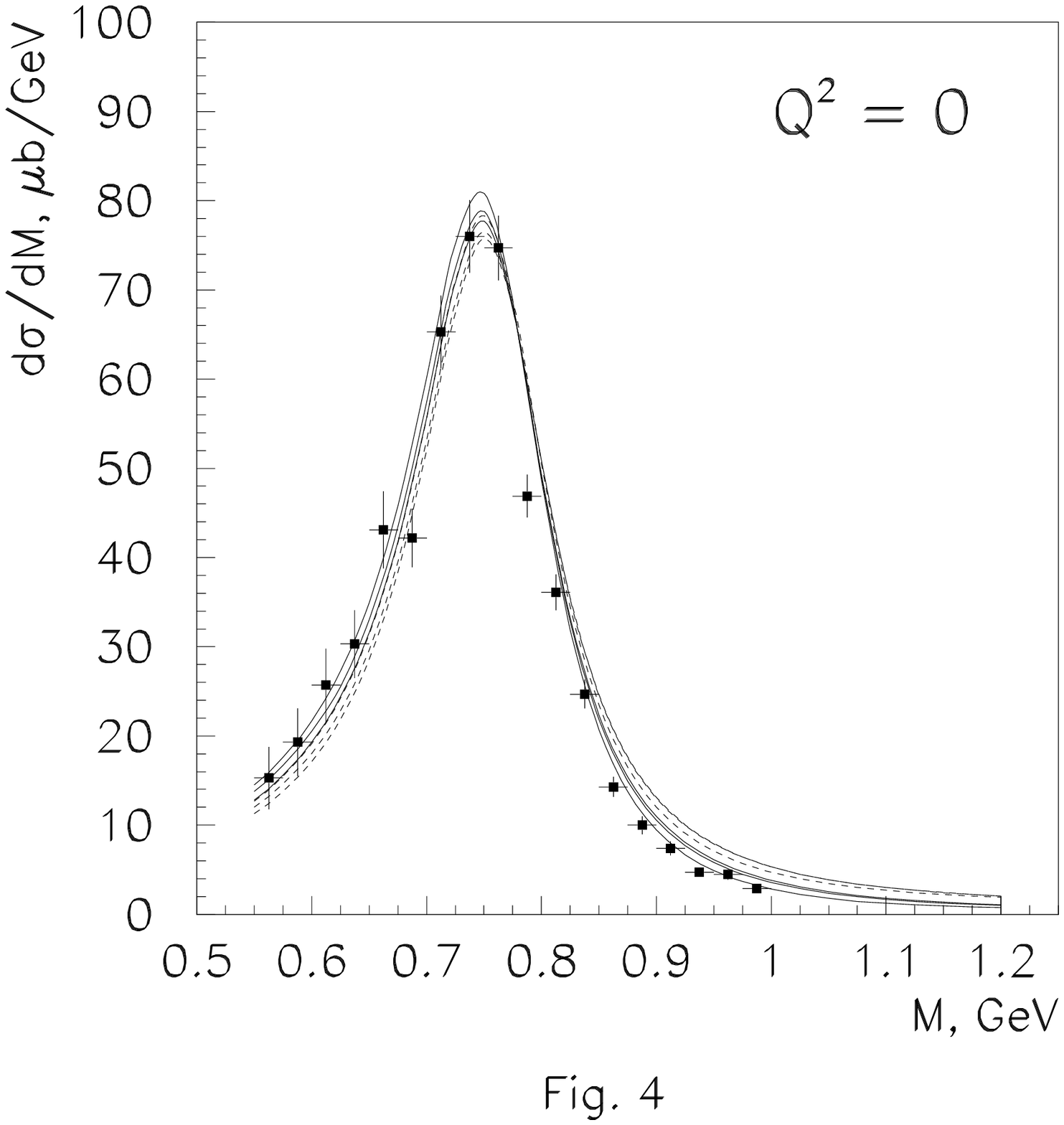,width=16cm}}
\end{figure}

\begin{figure}
\centerline{\epsfig{file=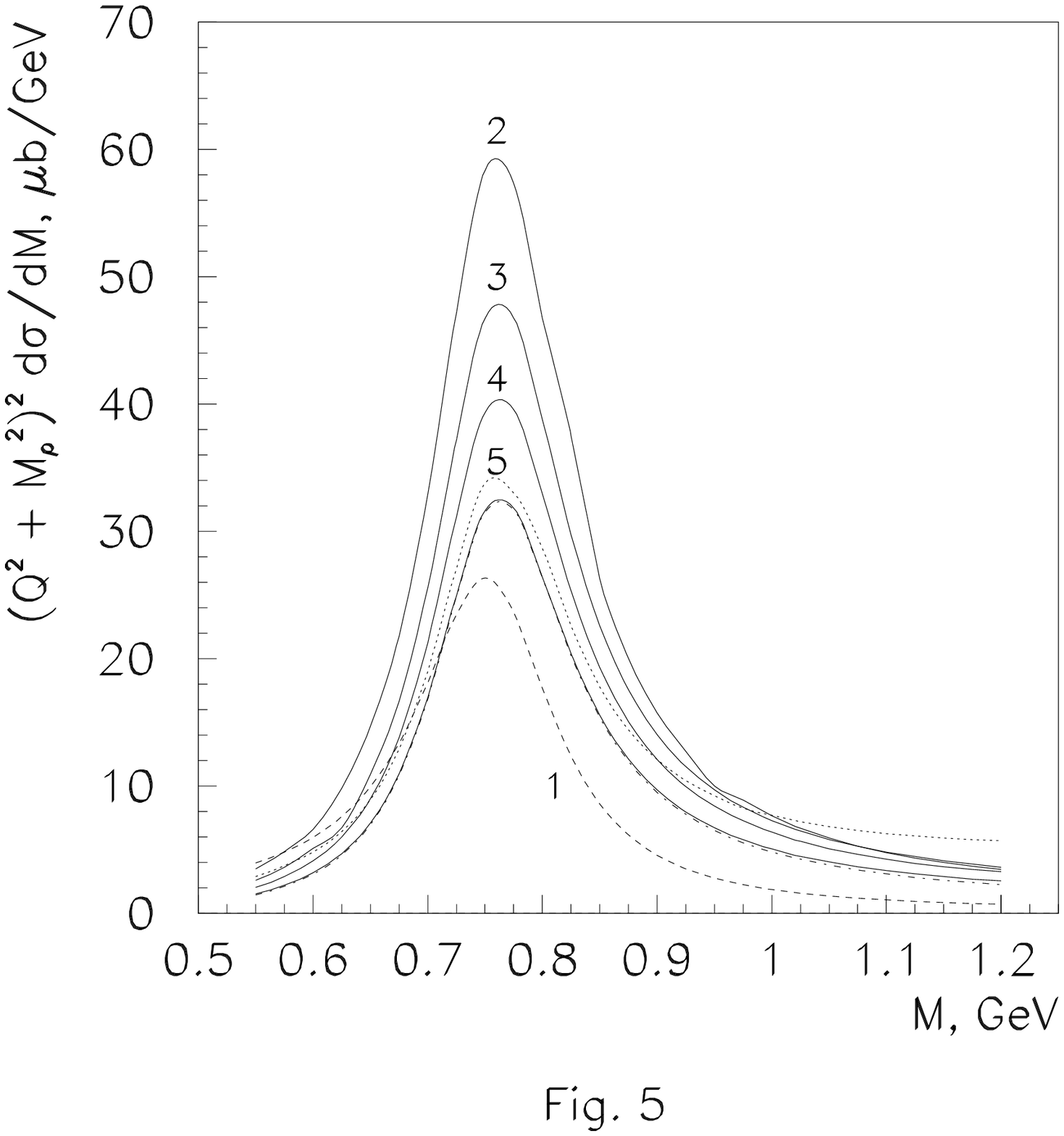,width=16cm}}
\end{figure}

\begin{figure}
\centerline{\epsfig{file=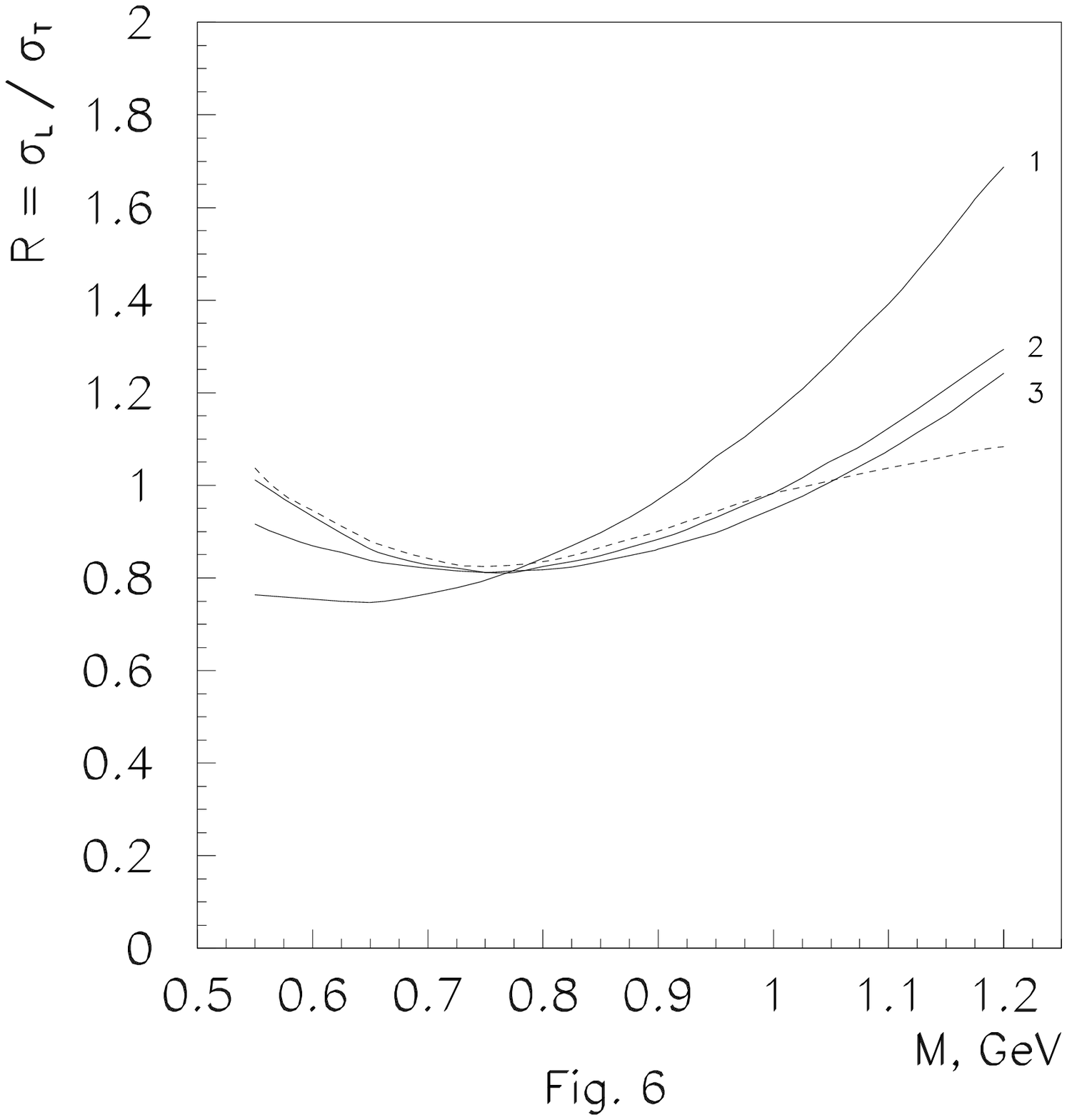,width=16cm}}
\end{figure}

\end{document}